\documentclass[twocolumn,preprintnumbers,amsmath,superscriptaddress,amssymb,aps]{revtex4-1}

\usepackage{graphicx}
\usepackage{dcolumn}
\usepackage{bm}
\usepackage{amsmath}
\usepackage{amssymb}
\usepackage{graphicx}
\usepackage{indentfirst}
\usepackage{booktabs}
\usepackage{multirow}
\usepackage{colortbl}

\selectfont

\begin{document}
\title{Origin of zigzag antiferromagnetic orders in XPS$_3$ (X= Fe, Ni) monolayers}

\author{Ping Li}
\thanks{These authors contributed equally to this work.}
\address{State Key Laboratory for Mechanical Behavior of Materials, Center for Spintronics and Quantum System, School of Materials Science and Engineering, Xi'an Jiaotong University, Xi'an, Shaanxi, 710049, China}
\address{State Key Laboratory for Surface Physics and Department of Physics, Fudan University, Shanghai, 200433, China}
\author{Xueyang Li}
\thanks{These authors contributed equally to this work.}
\address{Key Laboratory of Computational physical Sciences (Ministry of Education), Institute of Computational Physical Sciences, and Department of Physics, Fudan University, Shanghai, 200433, China}
\address{Shanghai Qi Zhi Institution, Shanghai, 200030, China}
\author{Junsheng Feng}
\address{School of Physics and Materials Engineering, Hefei Normal University, Hefei, 230601, China}
\author{Jinyang Ni}
\address{Key Laboratory of Computational physical Sciences (Ministry of Education), Institute of Computational Physical Sciences, and Department of Physics, Fudan University, Shanghai, 200433, China}
\address{Shanghai Qi Zhi Institution, Shanghai, 200030, China}
\author{Zhi-Xin Guo}
\email{zxguo08@xjtu.edu.cn}
\address{State Key Laboratory for Mechanical Behavior of Materials, Center for Spintronics and Quantum System, School of Materials Science and Engineering, Xi'an Jiaotong University, Xi'an, Shaanxi, 710049, China}
\author{Hongjun Xiang}
\email{hxiang@fudan.edu.cn}
\address{Key Laboratory of Computational physical Sciences (Ministry of Education), Institute of Computational Physical Sciences, and Department of Physics, Fudan University, Shanghai, 200433, China}
\address{Shanghai Qi Zhi Institution, Shanghai, 200030, China}
\address{Collaborative Innovation Center of Advanced Microstructures, Nanjing, 210093, China}

\date{\today}

\begin{abstract}
Recently, two monolayer magnetic materials, i.e., FePS$_3$ and NiPS$_3$, have been successfully fabricated. Despite that they have the same atomic structure, the two monolayers exhibit distinct magnetic properties. FePS$_3$ holds an out-of-plane zigzag antiferromagnetic (AFM-ZZ) structure, while NiPS$_3$ exhibits an in-plane AFM-ZZ structure. However, there is no theoretical model which can properly describe its magnetic ground state due to the lack of a full understanding of its magnetic interactions. Here, by combining the first-principles calculations and the newly developed machine learning method, we construct an exact spin Hamiltonian of the two magnetic materials.  Different from the previous studies which failed to fully consider the spin-orbit coupling effect, we find that the AFM-ZZ ground state in FePS$_3$ is stabilized by competing ferromagnetic nearest-neighbor and antiferromagnetic third nearest-neighbor exchange interactions, and combining single-ion anisotropy. Whereas, the often ignored nearest-neighbor biquadratic exchange is responsible for the in-plane AFM-ZZ ground state in NiPS$_3$. We additionally calculate spin-wave spectrum of AFM-ZZ structure in the two monolayers based on the exact spin Hamiltonian, which can be directly verified by the experimental investigation. Our work provides a theoretical framework for the origin of AFM-ZZ ground state in two-dimensional materials.
\end{abstract}

\maketitle
\section{Introduction}
Since the discovery of graphene, two-dimensional (2D) atomic crystals have seen a surge of interest due to their highly tunable physical properties and great potential in scalable device applications \cite{1,2,3,4,5,6}. The recent reports of ferromagnetic (FM) order in two different 2D crystals, Cr$_2$Ge$_2$Te$_6$ and CrI$_3$ \cite{7,8}, mark the beginning of a new chapter in the remarkable field of 2D materials. These discoveries significantly extend the list of electronically ordered 2D crystals, which includes superconductors \cite{9,10}, charge density wave materials \cite{11}, topological insulators \cite{12}, and ferroelectrics \cite{13,14}. The physical mechanisms of 2D FM materials has been described by the bilinear spin Hamiltonians (including Heisenberg symmetric exchange, Dzyaloshinskii-Moriya anti-symmetric exchange, anisotropic symmetric exchange, and single-ion anisotropy) \cite{15}.

Among 2D materials, transition-metal trichalcogenides XPS$_3$ (here we focus on X = Fe, Ni), are particularly interesting. They all have the same monoclinic structure with a space group $\emph{C2/m}$, where layers on the $\emph{ab}$ plane are coupled by a weak van der Waals force along the $\emph{c}$-axis \cite{16}. In these materials, the metal atoms are enclosed in octahedra formed by the sulfur atoms and there is a phosphorus doublet at the center of the honeycomb hexagons. Due to the slightly distorted octahedral crystal field, the  3$\emph{d$^6$}$ electrons of Fe hold the filled $\emph{e}$$_g$, $\emph{a}$$_{1g}$, $\emph{e}$$_{g}$' majority states and half-filled $\emph{e}$$_{g}$' minority states, and the 3$\emph{d$^8$}$ electrons of Ni hold the filled $\emph{a}$$_{1g}$,  e$_{g}$' majority and minority states and filled $\emph{e}$$_{g}$ minority states. This feature makes FePS$_3$ and NiPS$_3$  semiconducting materials with the magnetic moment of 4$\mu_B$ per Fe, and 2$\mu_B$ per Ni, respectively. Neutron scattering and Raman experiments reported that FePS$_3$ exhibits an Ising-type AFM-ZZ order down to the monolayer limit \cite{17,18,19,20}. Whereas, in-plane AFM-ZZ order was observed in  NiPS$_3$ \cite{21,22,23}. Theoretically, the origin of AFM-ZZ order in XPS$_3$ is still under debate. For example. Tae $\emph{et al.}$ reported that the dipolar anisotropy is essential to stabilize the AFM-ZZ state for TMPS$_3$ (TM = Mn, Fe, Ni) \cite{24}, while Mohammad $\emph{et al.}$ proposed that the orbital ordering induced by a variation of Fe-Fe pair distance is responsible for the AFM-ZZ order in FePS$_3$ \cite{25}. Therefore, a more systematical study on the spin Hamiltonian of 2D XPS$_3$ materials, which can exactly reveal the origin of their AFM-ZZ magnetic ground state, is highly desirable.

In this work, we explore the origin of AFM-ZZ ground state in 2D XPS$_3$, by constructing an exact spin Hamiltonian, which is realized by combining the first-principles calculations (DFT) \cite{26} and the newly developed machine learning method. We find that the AFM-ZZ ground state in FePS$_3$ and NiPS$_3$ is originated from different mechanisms. The AFM-ZZ order in FePS$_3$ is established by the competition of FM nearest-neighbor (NN) and AFM third NN exchange interactions between the Ising-like Fe spins, whereas the usually overlooked biquadratic exchange is the dominating factor for the AFM-ZZ order in NiPS$_3$. Moreover, the degenerate $\emph{d$_{xy}$}$ and $\emph{d$_{x^2-y^2}$}$ orbitals of Fe lead to a positive magnetocrystalline anisotropy energy (MAE) value, and thus an out-of-plane magnetism in FePS$_3$. Whereas, in the NiPS$_3$ the majority of $\emph{d}$ orbitals of Ni contribute to a negative MAE, leading to an in-plane magnetic ground state. Finally, we predict the spin-wave spectrum of FePS$_3$ and NiPS$_3$, which are expected to be observed in the experiments.

\section{COMPUTATIONAL METHODS}
Our first-principles calculations are performed based on the projected augmented-wave method \cite{27} encoded in the Vienna $Ab$ $initio$ simulation package (VASP) \cite{28}. Because the Perdew-Becke-Erzenh exchange-correlation functional is unable to give rise to correct $\emph{d}$-orbital occupied state of FePS$_3$, the local density approximation (LDA) exchange-correlation functional is used for the FePS$_3$ calculations \cite{29}. Nonetheless, the PBE exchange-correlation functional is adopted for the NiPS$_3$ calculations \cite{30}. The plane-wave basis with a kinetic energy cutoff of 500 eV is employed. To describe strongly correlated 3$\emph{d}$ electrons of Fe and Ni, the LDA+U and GGA+U methods are applied with the effective U value (U$_{eff}$ = U - J) of 4 eV, respectively. The spin-orbit coupling (SOC) effect is considered in the training set and testing set calculations of FePS$_3$, while it is not taken into account in the training set and testing set calculations of NiPS$_3$. This is because the double degenerate e$_{g}$' minority states are only filled with one electron for FePS$_3$, which is easily perturbed by orbital contributions. However, due to the filled $\emph{a}$$_{1g}$, $\emph{e}$$_{g}$' states and half-filled $\emph{e}$$_{g}$ states of Ni 3$\emph{d}$$^8$ electrons in NiPS$_3$, the SOC effect can be neglected. A vacuum of 20 $\rm \AA$ is set along the $\emph{c}$-axis, to avoid the interaction between the sheet and its periodic images. The convergence criteria of the total energy and the force are set to be 10$^{-6}$ eV and -0.01 eV/$\rm \AA$, respectively.

Spin exchange parameters are obtained by combing the Machine Learning Method for Constructing Hamiltonian (MLMCH) \cite{26,31}, four-state method \cite{15,32}, and modified four-state mapping method \cite{32}. By applying machine learning approaches and statistical analysis, MLMCH is able to find out the most important interaction terms among thousands of candidate terms efficiently and correctly. The truncation distances of 2nd and 4th order terms are both set to 20 Bohr. After symmetry analysis, the number of possible nonequivalent parameters (p$_{max}$) are 24 for FePS$_3$ and 76 for NiPS$_3$ (including a constant term), respectively. The training set and the testing set contain 150 and 50 sets of data, respectively. For four-state method and MLMCH, we use 3$\times$3$\times$1 supercell of monolayer XPS$_3$ to extract the related magnetic parameters. In our parallel tempering Monte Carlo (PTMC) simulations of spin Hamiltonian \cite{33,34} with the PASP package \cite{35}, a 36$\times$36$\times$1 supercell of the unit cell is adopted for monolayer XPS$_3$. Similar results are obtained with larger supercells (48$\times$48$\times$1) to estimate the magnetic critical temperature. The number of replicas is set to 80. The spin-wave spectrum is calculated within the linear spin wave theory (LSWT) using the SpinW software package \cite{36}.

\section{RESULTS AND DISCUSSION }	
\subsection{Spin Hamiltonian }

\begin{table*}[htb]
\caption{
The monolayer XPS$_3$ lattice constants \emph{a} ($\rm \AA$) are optimized. The bilinear exchange interactions (in meV) and NN biquadratic interaction (in meV) were calculated with MLMCH method (or four-state method) for the monolayer XPS$_3$. The SOC effect is included in FePS$_3$.}
\begin{tabular*}{0.6\linewidth}{cccccccc}
	\hline
	              & \emph{a}  & \emph{J$_1$ }   & \emph{J$_2$}  & \emph{J$_3$}  & \emph{K}          & \emph{A$_z$}   \\
	\hline
	FePS$_3$      & 5.82      & -2.20(-2.10)    & NA(0.10)       & 2.07(2.18)   & -2.69(NA)         &  5.76(4.88) \\
	NiPS$_3$      & 5.86      & -3.60(-3.38)    & -0.60(-0.64)  & 15.65(15.79)  & -1.56(-1.66)      &  NA(-1.25)   \\
	\hline
\end{tabular*}
\end{table*}	

We first calculate the relative energies for four possible magnetic configurations in a 2$\times$2$\times$1 supercell (see Fig. 1), namely, the FM, N$\acute{e}$el antiferromagnetic (AFM-N), AFM-ZZ, stripy antiferromagnetic (AFM-ST) structures, using two different procedures \cite{37}. One is the structure optimized with the FM spin order (see Fig. 1), the other uses the structure optimized with the FM, AFM-N, AFM-ZZ, and AFM-ST spin order, respectively (see Fig. S1). As shown in Fig. 1 and Fig. S1, the two procedures give similar energetics with the AFM-ZZ order being most stable, which agrees well with previous studies \cite{24,25}. This result also indicates that spin-lattice coupling can be neglected. Hereafter, we only consider the spin degrees of freedom.

We then construct the exact spin Hamiltonian based on the MLMCH calculations. After extensive calculations, we obtain several most significant interaction terms among thousands of candidates for the spin Hamiltonian. It is found that the spin Hamiltonian of XPS$_3$ monolayers has a general form
\begin{equation}
\begin{split}
H = & \sum_{\langle i,j\rangle}[ J_{1}S_i \cdot S_j + K(S_i \cdot S_j)^2] + \sum_{\langle i,l\rangle} J_{2}S_i \cdot S_l \\ & + \sum_{\langle i,k\rangle} J_{3}S_i \cdot S_k - \sum_{i} A_zS_{iz}^2,
\end{split}
\end{equation}
where $\emph{J$_1$}$, $\emph{K}$, $\emph{J$_2$}$, $\emph{J$_3$}$, and $\emph{A$_z$}$ are first NN Heisenberg exchange parameter, first NN biquadratic exchange parameter, second NN Heisenberg exchange parameter, third NN Heisenberg exchange parameter, and single-ion anisotropy parameter, respectively. The negative and positive values represent FM and AFM interactions for Heisenberg interaction, respectively. As shown in Table I, in the XPS$_3$ monolayers the NN FM exchange interactions $\emph{J$_1$}$ and third NN AFM interactions $\emph{J$_3$}$ are very strong, whereas the second NN interactions $\emph{J$_2$}$ can be neglected. Note that the unusually large $\emph{J$_3$}$ in FePS$_3$ can be obtained via DFT calculations provided that the SOC effect is additionally considered in the atomic-structure optimization procedure, which was ignored in previous studies \cite{24,25,38}. It is also noticed that both FePS$_3$ and NiPS$_3$ have significant biquadratic interactions.

On the other hand, in order to confirm the above MLMCH results, we further calculate the interaction parameters by means of the four-state method \cite{15,29} using a 3$\times$3$\times$1 supercell. As shown in Table I, the spin Hamiltonian parameters obtained by the two methods are consistent.

\begin{figure}[htb]
\begin{center}
\includegraphics[angle=0,width=1.0\linewidth]{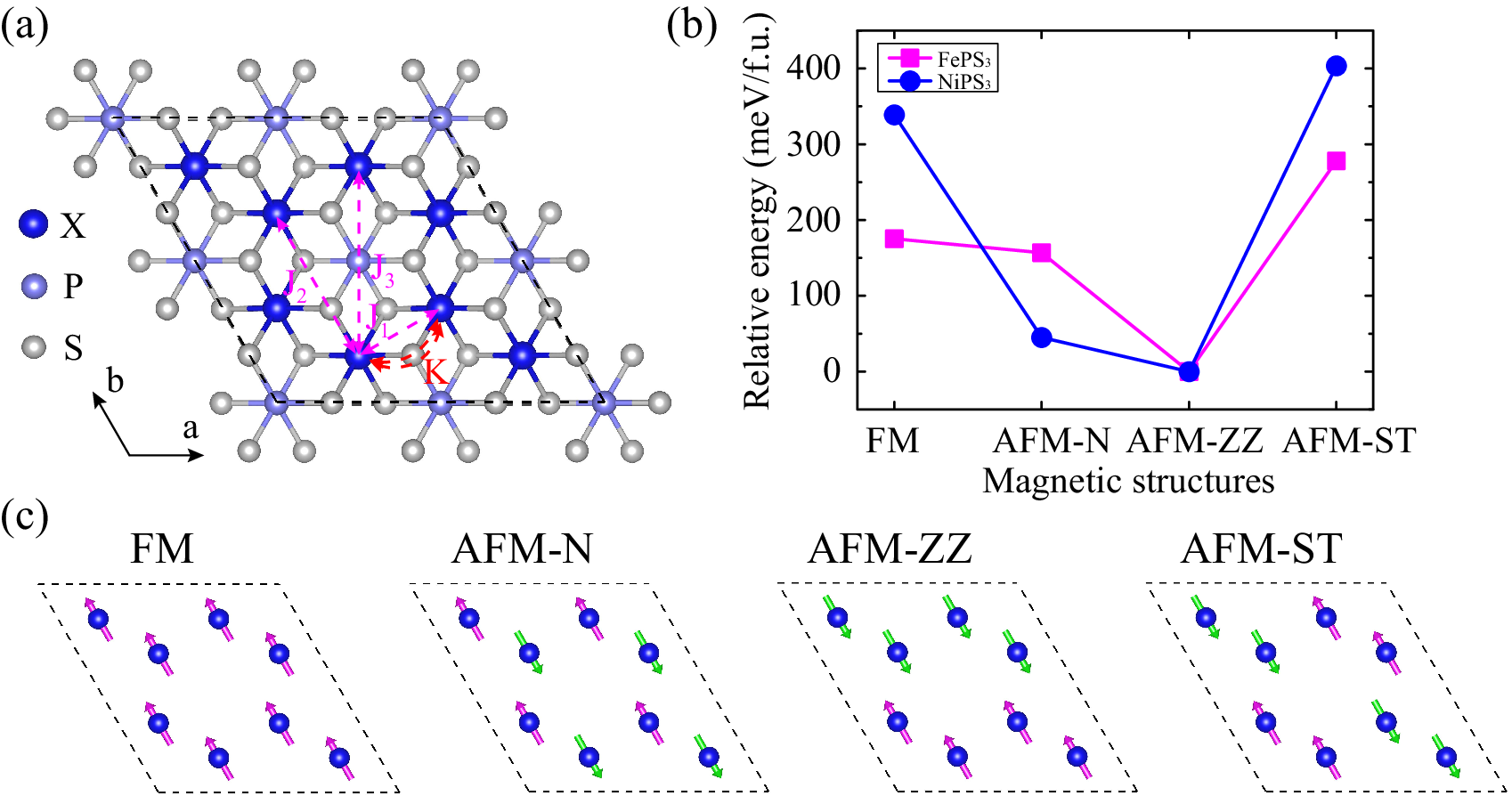}
\caption{(a) The crystal structure of monolayer XPS$_3$, along with magnetic exchange paths [i.e., first-neighbor (\emph{J$_1$}), first-neighbor biquadratic ($\emph{K}$), second-neighbor (\emph{J$_2$}), and third-neighbor (\emph{J$_3$})]. (b) Calculated relative energies for various magnetic structures of monolayer FePS$_3$ by LDA+U+SOC (NiPS$_3$ by GGA+U) method using the structure optimized with the FM order. Here, the AFM-ZZ state is chosen as the energy reference. (c) Schematic top view of various magnetic structures only containing magnetic atoms.}
\end{center}
\end{figure}

We further discuss the underlying mechanism for the obtained large exchange interactions in XPS$_3$ monolayers. The FM spin exchange between the nearest neighbor pair competes with AFM third-neighbor exchange interactions. Our structure analysis shows that the bond angle of X-S-X is close to 90$^{\circ}$, suggesting that the super-exchange leads to the sizable first NN FM exchange interaction. On the other hand, the indirect super-superexchange interaction result in the unusually large third NN AFM exchange interaction. As for the biquadratic interaction $\emph{K}$, the interesting phenomenon is that FePS$_3$ and NiPS$_3$ both have a large $\emph{K}$.

Note that the SOC effects are different in the two materials. In FePS$_3$, only one electron is filled in the double degenerate e$_{g}$' minority state, which results in a strong SOC effect. Whereas, the SOC is much weaker in NiPS$_3$ owing to the fully filled double degenerate states. We additionally show in Fig. S2 the relative energies of various magnetic structures of  FePS$_3$ calculated without SOC. It is found that although the magnetic ground state is still AFM-ZZ, the obtained exchange interactions are completely different from that obtained with SOC (Table SI). More importantly, the AFM-ZZ ground state cannot be obtained when such exchange interaction parameters are used in the Monte Carlo simulation (see Fig. S3(a)). This result shows that the SOC effect cannot be ignored when investigating the magnetic properties of FePS$_3$. It is also noticed that FePS$_3$ has a robust out-of-plane magnetization, but NiPS$_3$ exhibits an easy-plane magnetization. Considering the magnetic moment switching from the out-of-plane [001] axis to the in-plane [100] axis, the energy difference can be defined as MAE, i.e., MAE = \emph{E$_{100}$ }- \emph{E$_{001}$}. Generally, MAE is induced by the crystal field splitting and SOC effect. In order to reveal the physical mechanism of MAE difference between FePS$_3$ and NiPS$_3$, we perform  analysis based on the second-order perturbation theory \cite{39,40,41}, where the MAE can be approximately described by
\begin{equation}
\begin{split}
\Delta E^{sl} & = E_x^{sl} - E_z^{sl} \\ & = \xi^2\sum_{o}\sum_{u} \frac{|\langle o| \textbf{S$_x$} \cdot \textbf{L$_x$} |u \rangle|^2 - |\langle o| \textbf{S$_z$} \cdot \textbf{L$_z$} |u \rangle|^2}{\varepsilon_{o} - \varepsilon_{u}},
\end{split}
\end{equation}
with $\emph{E$_x$}$ and $\emph{E$_z$}$ being the SOC energies for the x-axis and z-axis magnetization directions, respectively. $|o\rangle $ and $|u\rangle$ denote the occupied and unoccupied states, respectively. According to Eq. (2), MAE is determined by matrix elements of the spin-orbital interaction between occupied and unoccupied states.

\begin{figure}[htb]
\begin{center}
\includegraphics[angle=0,width=1.0\linewidth]{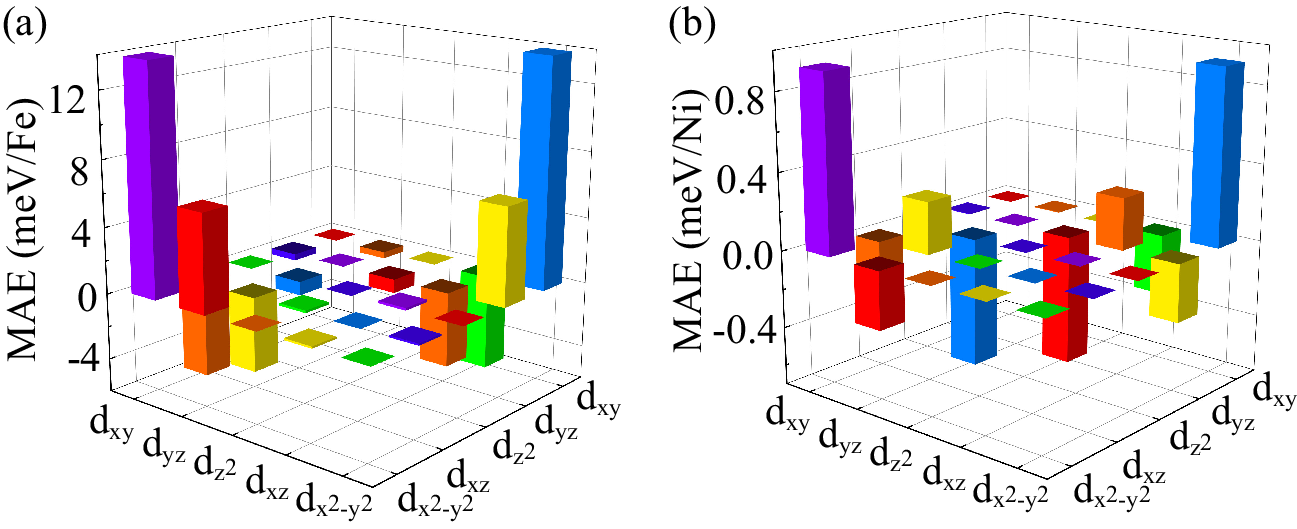}
\caption{Atomic orbital-resolved magnetocrystalline anisotropy energy (MAE) difference of (a) Fe atom in monolayer FePS$_3$, and (b) Ni atom in monolayer NiPS$_3$.}
\end{center}
\end{figure}

Fig. 2 and Fig. S4 show the calculated orbital-resolved MAE, where positive and negative values of MAE denotes the out-of-plane and in-plane magnetizations, respectively. Clearly, the MAE mainly originates from Fe (in FePS$_3$) and Ni (in NiPS$_3$) atoms, while those from P and S atoms have  minor contributions. In the FePS$_3$, the $\emph{d$_{xy}$}$ and $\emph{d$_{x^2-y^2}$}$, $\emph{d$_{yz}$}$ and $\emph{d$_{x^2-y^2}$}$ orbitals of Fe mainly contribute to the positive MAE values. Whereas, the $\emph{d$_{xy}$}$ and $\emph{d$_{xz}$}$, $\emph{d$_{xz}$}$ and $\emph{d$_{yz}$}$ orbitals contribute to the negative MAE values. This result strongly indicates that the FePS$_3$ monolayer possesses an out-of-plane magnetization. In the NiPS$_3$,  on the other hand, the  $\emph{d$_{xy}$}$ and $\emph{d$_{x^2-y^2}$}$, $\emph{d$_{xz}$}$ and $\emph{d$_{yz}$}$ orbitals of Ni contribute to positive MAE values, while the remaining $\emph{d}$ orbitals give rise to negative MAE values. As a result of competition, the NiPS$_3$ exhibits an in-plane MAE.

The MAE can be further understand from the SOC formula. Generally, the $\hat{H}_{SO}$ = $\lambda$$\hat{S}$$\cdot$$\hat{L}$ term can be written as \cite{15,42,43}
\begin{equation}
\begin{split}
\hat{H}_{SO} & = \lambda \hat{S}_{z'}(\hat{L}_zcos\theta + \frac{1}{2}\hat{L}_+e^{-i\varphi}sin\theta + \frac{1}{2}\hat{L}_-e^{i\varphi}sin\theta) \\ & + \frac{\lambda}{2}\hat{S}_{+'}(-\hat{L}_zsin\theta - \hat{L}_+e^{-i\varphi}sin^2\frac{\theta}{2} + \hat{L}_-e^{i\varphi}cos^2\frac{\theta}{2}) \\ & + \frac{\lambda}{2}\hat{S}_{\_'}(-\hat{L}_zsin\theta + \hat{L}_+e^{-i\varphi}cos^2\frac{\theta}{2} - \hat{L}_-e^{i\varphi}sin^2\frac{\theta}{2}).
\end{split}
\end{equation}
To a qualitative discussion of spin orientation, the SOC Hamiltonian $\hat{H}$$_{SO}$ can be rewritten as
\begin{equation}
\hat{H}_{SO} = \hat{H}^0_{SO} + \hat{H}^1_{SO},
\end{equation}
where $\hat{H}$$^0_{SO}$ is the "spin-conserving" term,
\begin{equation}
\hat{H}^0_{SO} = \lambda \hat{S}_{z'}(\hat{L}_zcos\theta + \frac{1}{2}\hat{L}_+e^{-i\varphi}sin\theta + \frac{1}{2}\hat{L}_-e^{i\varphi}sin\theta),
\end{equation}
and $\hat{H}$$^1_{SO}$ is the "spin-non-conserving" term \cite{15}.
Since the lowest energy gap between the occupied and unoccupied levels of the XPS$_3$ occurs at a spin down state, so only the "spin-conserving" term is considered. For the FePS$_3$, the Fe$^{2+}$ ($\emph{d}$$^6$) ion has the d-state splitting pattern (e$_{g'}$ $\uparrow$)$^2$ $<$ (a$_{1g}$ $\uparrow$)$^1$ $<$ (e$_{g}$ $\uparrow$)$^2$ $<$ (e$_{g'}$ $\downarrow$)$^1$ $<$ (a$_{1g}$ $\downarrow$)$^0$ $<$ (e$_{g}$ $\downarrow$)$^0$. The lowest energy gap between the occupied and unoccupied levels occurs in (e$_{g'}$ $\downarrow$). Since their $\emph{m}$ (magnetic quantum number) values are the same $\Delta m$ = 0, they can interact when the spin direction is parallel to the orbital z-axis. In other words, the preferred spin direction is parallel to the orbital z-axis. We further consider the Ni$^{2+}$ ($\emph{d}$$^8$) ion of the NiPS$_3$ with the $\emph{d}$-electron configuration (e$_{g'}$ $\uparrow$)$^2$ $<$ (a$_{1g}$ $\uparrow$)$^1$ $<$ (e$_{g}$ $\uparrow$)$^2$ $<$ (e$_{g'}$ $\downarrow$)$^2$ $<$ (a$_{1g}$ $\downarrow$)$^1$ $<$ (e$_{g}$ $\downarrow$)$^0$. Therefore, the lowest energy gap occurs for the energy level difference of a$_{1g}$ $\downarrow$ and e$_{g}$ $\downarrow$, because these two orbitals cannot interact due to the nonzero $\Delta m$ = 2. The next lowest energy gap occurs for the e$_{g'}$ $\downarrow$ and e$_{g}$ $\downarrow$ levels, they can interact because their m values differ by $\pm 1$. Namely, the preferred spin orientation is perpendicular to the z-axis. Consequently, both above two theories give rise to the same results, which perfectly reveal the physical origins of MAE in XPS$_3$.

\subsection{Microscopic mechanisms of AFM-ZZ state in XPS$_3$ (X= Fe, Ni) monolayers}
In the above study, we have obtained the spin Hamiltonian by the MLMCH calculations. To determine the magnetic ground state of XPS$_3$ monolayers, we carried out PTMC simulations with spin Hamiltonian. More interestingly, although their spin Hamiltonian differ, the AFM-ZZ magnetic ground state are both obtained by the PTMC simulations (see Fig. S3 and Fig. S5). In order to have a comprehensive understanding on the microscopic mechanism of magnetic ground state of XPS$_3$ monolayers, we additionally calculate the phase diagram using the obtained spin Hamiltonian model (Fig. 3) and the PTMC simulations \cite{33,34}.

\begin{figure}[htb]
\begin{center}
\includegraphics[angle=0,width=1.0\linewidth]{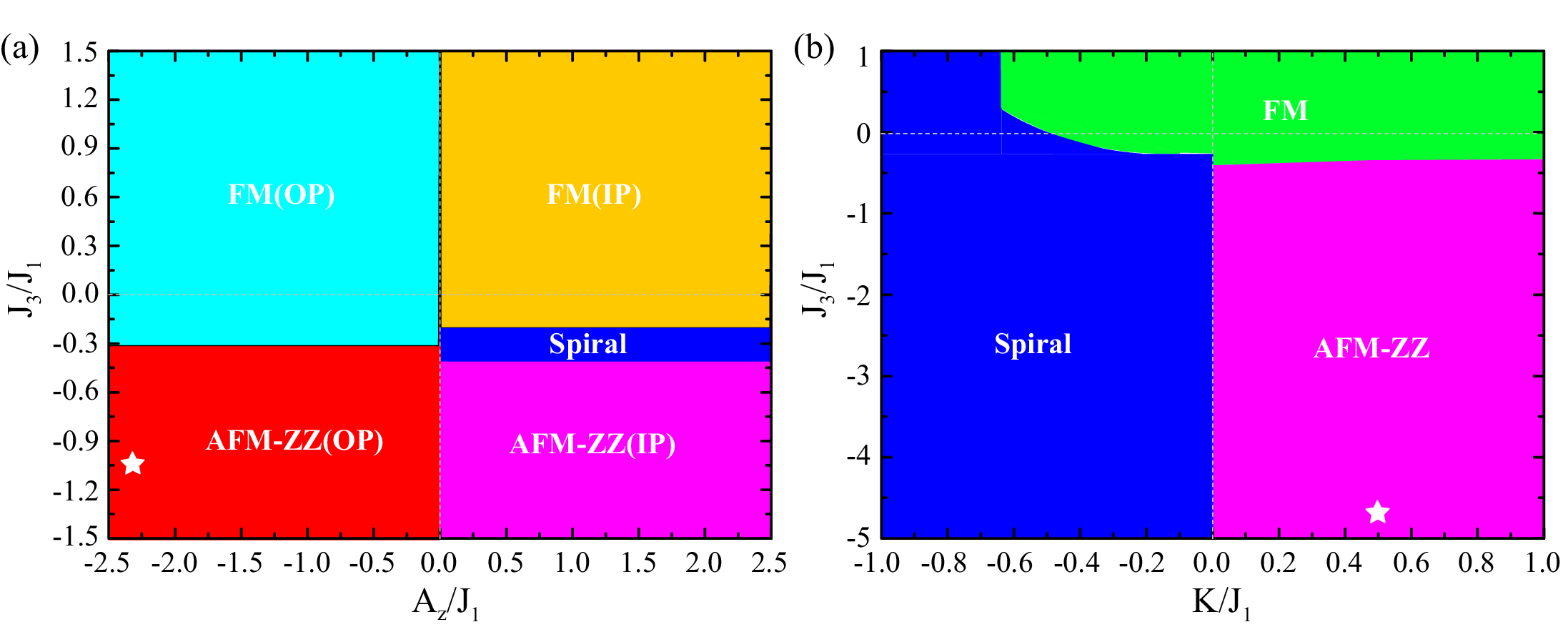}
\caption{Phase diagrams of the \emph{J$_1$}-\emph{J$_3$}-\emph{A$_z$} model for (a) FePS$_3$, and the \emph{J$_1$}-\emph{K}-\emph{J$_3$} model for (b) NiPS$_3$. All simulations are based on \emph{J$_1$} $<$ 0. The white pentagram represents the magnetic ground state. }
\end{center}
\end{figure}

In the FePS$_3$, the $\emph{J$_1$-J$_3$-A$_z$}$ terms are considered in the simulations. The $\emph{J$_1$}$ and $\emph{J$_3$}$ terms dominate  magnetically ordered states, and the $\emph{A$_z$}$ term determines the direction of magnetization. As shown in Fig. 3(a), when $\emph{A$_z$/J$_1$}$ $<$ 0, FePS$_3$ presents an out-of-plane FM state with $\emph{J$_3$/J$_1$}$ $>$ -0.33. Whereas, the out-of-plane AFM-ZZ state appears when increasing the AFM interaction ($\emph{J$_3$/J$_1$}$ $<$ -0.33). On the other hand,  in the region $\emph{A$_z$/J$_1$}$ $>$ 0, there are two transition points, i.e., the in-plane FM to spiral state and the spiral to in-plane AFM-ZZ state. Such two phase transitions occur around $\emph{J$_3$/J$_1$}$ $\simeq$ -0.20, and -0.40, respectively. We additionally explore the effect of biquadratic interaction $\emph{K}$ on the magnetic ground state of FePS$_3$, by changing $\emph{K}$ from -2.69 meV to 2.69 meV and keeping $\emph{J$_1$}$ = -2.20 meV, $\emph{J$_1$}$ = 2.07 meV, and $\emph{A$_z$}$ = 5.76 meV. The calculated results show that the magnetic ground state keeps AFM-ZZ when $\emph{K}$ $<$ 0 meV, while, it will become spiral state if $\emph{K}$ $>$ 0 meV. Hence, negative biquadratic interaction mainly has an effect of keeping the magnetic ground state collinear in FePS$_3$.

As for the NiPS$_3$, the $\emph{J$_1$-K-J$_3$}$ model is adopted, due to the significant biquadratic interaction. As shown in Fig. 3(b), in case of $\emph{K/J$_1$}$ $<$ 0, the FM state exists only in a small region, and the remaining regions are spiral states. When $\emph{K/J$_1$}$ $>$ 0, the FM order exists with $\emph{J$_3$/J$_1$}$ $>$ -0.33. There is also a  transition from FM to AFM-ZZ state with $\emph{J$_3$/J$_1$}$ $<$ -0.33. The above results show that the third NN AFM Heisenberg interaction $\emph{J$_3$}$ and first NN FM biquadratic interaction $\emph{K}$ are crucial factors for the formation of AFM-ZZ order.

\subsection{Spin-wave spectrum}
\begin{figure}
\begin{center}
\includegraphics[angle=0,width=1.0\linewidth]{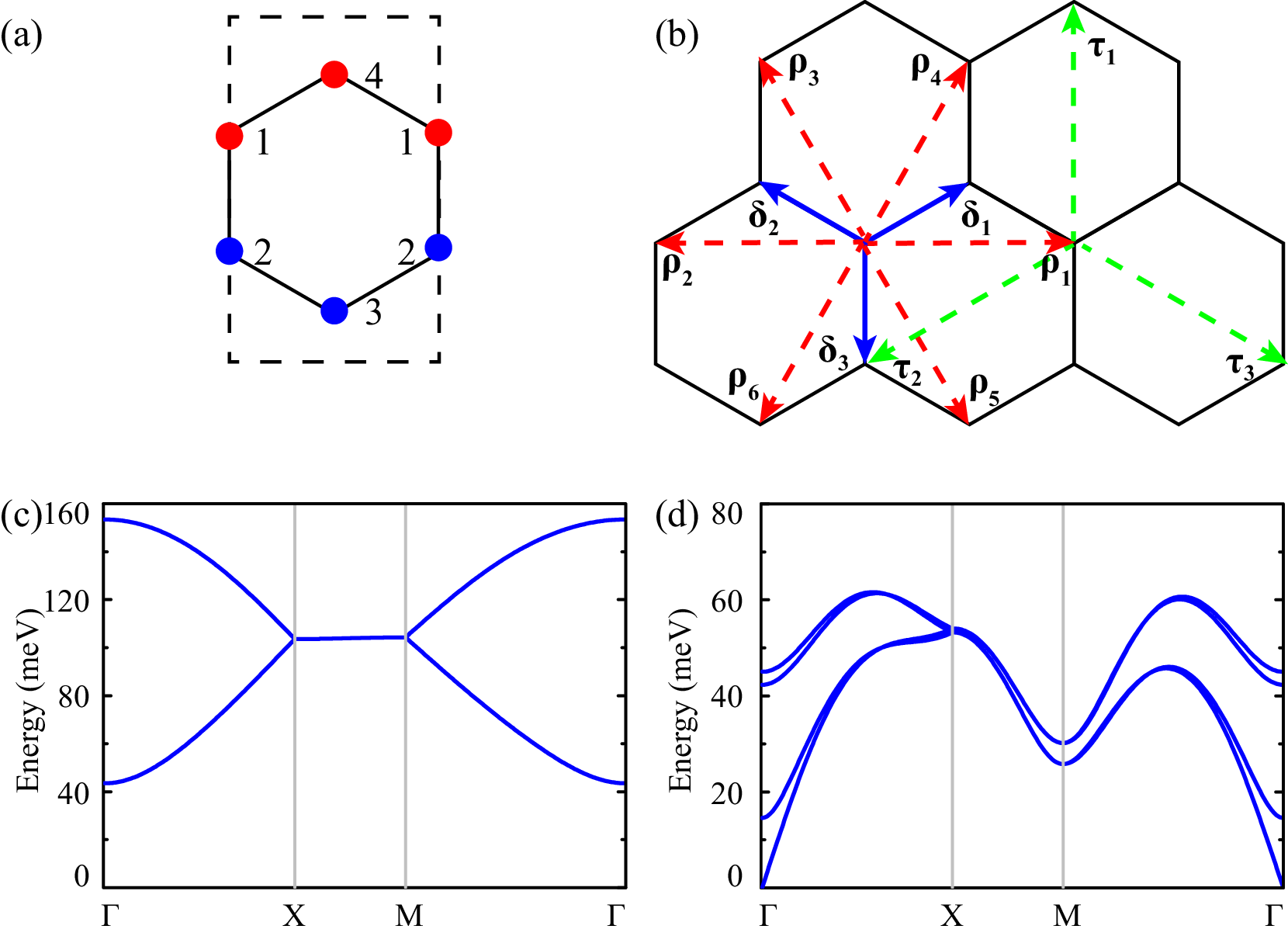}
\caption{(a) The sub-lattices inside magnetic unit cell. The spin up and spin down X (X = Fe, Ni) atoms are depicted by the red and blue circles, respectively. (b) The $\delta$, $\rho$, and $\tau$ are vectors joining nearest, second nearest and third nearest neighboring X atoms. Spin-wave spectrum of (c) FePS$_3$ and (d) NiPS$_3$ calculated using the LSWT method. The adopted magnetic interaction parameters are listed in Table I.}
\end{center}
\end{figure}

The obtained spin Hamiltonian allows us to make an accurate prediction on the spin-wave spectrum of XPS$_3$ monolayers. The spin-wave spectrum can be calculated by using the LSWT method \cite{36,46}, where the linearized Holstein-Primakoff transformation is adopted,
\begin{equation}\label{hp}
\begin{split}
\hat{S}_i^+ &= \sqrt{\frac{S}{2}}(\hat{b}_i^{\dag} + \hat{b}_i)    \\
\hat{S}_i^- &= i\sqrt{\frac{S}{2}}(\hat{b}_i^{\dag} - \hat{b}_i)    \\
\hat{S}_i^z &= S - \hat{b}_i^{\dag} \hat{b}_i .   \\
\end{split}
\end{equation}
Here $\hat{b}_i^{\dag}$ and $\hat{b}_i$ are the bosonic creation and annihilation operators on the site $i$, respectively. The representation of the spin Hamiltonian Eq. (\ref{hp}) in terms of this bosonic operators leads to a complicated non-linear Hamiltonian. In the spin-wave linear approximation, the higher order terms are neglected.

In the zigzag ordered materials, it has a doubled unit cell consisting of four lattice sites and a spin configuration, as shown in Fig. 4(a). For the FePS$_3$, the effective  Hamiltonian in momentum space is written as (for derivation of the Hamiltonian, see the Supplemental Material)
\begin{widetext}
\begin{equation}
  \hat{\mathrm{H}}(\mathbf{k})  = \left(
  \begin{array}{cccccccc}
  f_1(\mathbf{k})      & f_2(\mathbf{k})     & 0          & 0      & 0          & 0          & f_3(\mathbf{k})     & f_4(\mathbf{k}) \\
  f_2^{*}(\mathbf{k})  & f_1(\mathbf{k})     & 0          & 0      & 0          & 0          & f_4^{*}(\mathbf{k}) & f_3(\mathbf{k}) \\
  0           & 0          & f_1(\mathbf{k})     & f_2(\mathbf{k}) & f_3^{*}(\mathbf{k}) & f_4(\mathbf{k})    & 0          & 0 \\
  0           & 0          & f_2^{*}(\mathbf{k}) & f_1(\mathbf{k}) & f_4^{*}(\mathbf{k}) & f_3^{*}(\mathbf{k}) & 0          & 0 \\
  0           & 0          & f_3(\mathbf{k})     & f_4(\mathbf{k}) & f_1^{*}(\mathbf{k}) & f_2(\mathbf{k})     & 0          & 0 \\
  0           & 0          & f_4^{*}(\mathbf{k}) & f_3(\mathbf{k}) & f_2^{*}(\mathbf{k}) & f_1^{*}(\mathbf{k}) & 0          & 0 \\
  f_3^{*}(\mathbf{k})  & f_4(\mathbf{k})     & 0          & 0      & 0          & 0          & f_1^{*}(\mathbf{k}) & f_2(\mathbf{k}) \\
  f_4^{*}(\mathbf{k})  & f_3^{*}(\mathbf{k}) & 0          & 0      & 0          & 0          & f_2^{*}(\mathbf{k}) & f_1^{*}(\mathbf{k}) \\
  \end{array}
  \right)
\end{equation}
\end{widetext}
where $f_1(\mathbf{k}) = -J_{1}S - 6KS^3+ (2 + \xi_{1\mathbf{k}})J_{2}S + 3J_{3}S + A_{z}S$, $f_2(\mathbf{k}) = (J_{1}S + 2KS^3)\gamma_{1\mathbf{k}}$, $f_3(\mathbf{k}) = J_{2}S\xi_{2\mathbf{k}}$, and $f_4(\mathbf{k}) = (J_{1}S - 2KS^3)\gamma_{2\mathbf{k}} + J_{3}S\gamma_{\mathbf{k}}$. The explicit form of the structure factors $\gamma_{\mathbf{k}},\gamma_{1\mathbf{k}},\gamma_{2\mathbf{k}},\xi_{1\mathbf{k}},\xi_{1\mathbf{k}}$ are presented in Supplemental Material. Following, we calculate the magnon band structure of FePS$_3$ using the parameters obtained by the MLMCH method, as shown in Fig. 4(c). It shows that the out-of-plane single-ion anisotropy will open a spin-wave gap 43.50 meV in FePS$_3$. Moreover, we find that the spectrum is fourfold degenerate along the X - M line, which splits into two doubly degenerate bands on the $\Gamma$ - X and M - $\Gamma$ lines due to the out-of-plane anisotropy. On the X - M line, (T$\bar{M}$$_y$)$^2$ = -1, indicating a Kramers degeneracy. On the other hand, T$\bar{M}$$_y$ commutes with S$_x$, indicating the Kramers pair is within the same eigenspace of S$_x$. In addition, the spin-wave spectrum has two branches at the $\Gamma$ point. Based on the magnetic point group, the lower energy magnon is the Raman-active A$_g$ mode, while the higher energy magnon belongs to representation B$_u$\cite{47}. This result is completely consistent with previous experimental reports \cite{47}.

For the NiPS$_3$, although it has similar structures as FePS$_3$, the spin direction of magnetic ground state lies in the $ab$ plane, different from that of FePS$_3$. Hence, NiPS$_3$ has a different Hamiltonian (for derivation of the Hamiltonian, see the Supplemental Material) that can be written as
\begin{widetext}
\begin{equation}
  \hat{\mathrm{H}}(\mathbf{k})  = \left(
  \begin{array}{cccccccc}
  g_1(\mathbf{k})      & g_2(\mathbf{k})     & 0           & 0      & g_5(\mathbf{k})     & 0            & g_3(\mathbf{k})     & g_4(\mathbf{k}) \\
  g_2^{*}(\mathbf{k})  & g_1(\mathbf{k})     & 0           & 0      & 0          & g_5(\mathbf{k})       & g_4^{*}(\mathbf{k}) & g_3(\mathbf{k}) \\
  0           & 0          & g_1(\mathbf{k})      & g_2(\mathbf{k}) & g_3^{*}(\mathbf{k}) & g_4(\mathbf{k})       & g_5(\mathbf{k})     & 0 \\
  0           & 0          & g_2^{*}(\mathbf{k}) & g_1(\mathbf{k}) & g_4^{*}(\mathbf{k}) & g_3^{*}(\mathbf{k})   & 0          & g_5(\mathbf{k}) \\
  g_5^{*}(\mathbf{k})  & 0          & g_3(\mathbf{k})      & g_4(\mathbf{k}) & g_1^{*}(\mathbf{k}) & g_2(\mathbf{k})       & 0          & 0 \\
  0           & g_5^{*}(\mathbf{k}) & g_4^{*}(\mathbf{k})  & g_3(\mathbf{k}) & g_2^{*}(\mathbf{k}) & g_1^{*}(\mathbf{k})   & 0          & 0 \\
  g_3^{*}(\mathbf{k})  & g_4(\mathbf{k})     & g_5^{*}(\mathbf{k})  & 0      & 0          & 0            & g_1^{*}(\mathbf{k}) & g_2(\mathbf{k}) \\
  g_4^{*}(\mathbf{k})  & g_3^{*}(\mathbf{k}) & 0           & g_5^{*}(\mathbf{k}) & 0      & 0            & g_2^{*}(\mathbf{k}) & g_1^{*}(\mathbf{k}) \\
  \end{array}
  \right)
\end{equation}
\end{widetext}
where $g_1(\mathbf{k}) = -J_{1} - 6 K S^3 + (2 + \xi_{1\mathbf{k}})J_{2}S + 3J_{3}S - \frac{1}{2} A_{z}S$, $g_2(\mathbf{k}) = (J_{1}S + 2KS^3)\gamma_{1\mathbf{k}}$, $g_3(\mathbf{k}) = -J_{2}S\xi_{2\mathbf{k}}$, $g_4(k) = (-J_{1}S + 2KS^3)\gamma_{2\mathbf{k}} - J_{3}S\gamma_{\mathbf{\mathbf{k}}}$, and $g_5(\mathbf{k}) = - \frac{1}{2} A_{z}S$. The explicit form of the structure factors $\gamma_{\mathbf{k}},\gamma_{1\mathbf{k}},\gamma_{2\mathbf{k}},\xi_{1\mathbf{k}},\xi_{1\mathbf{k}}$ are listed in Supplemental Material. As shown in Fig. 4(d), we employ the Heisenberg exchange parameter by the MLMCH method and single-ion anisotropy parameter by the four-state method to calculate the magnon band structure of NiPS$_3$. It is seen that, when the direction of easy magnetization is turned into in-plane, the spin-wave gap will disappear in NiPS$_3$. Moreover, the double degeneracy of spin-wave spectrum is broken, and simultaneously, an hourglass band appears in the $\Gamma$ - X direction of the high symmetry line. It originates from the fact that S$_x$ is no longer a good quantum number. The calculated spin-wave spectrum provides a theoretical guidance for the experimental investigation on the magnetic dynamics of 2D XPS$_3$.

\section{CONCLUSION}
In summary, to understand the microscopic mechanisms of the AFM-ZZ ground state in XPS$_3$ (X = Fe, Ni) monolayer system, we construct the spin Hamiltonian by combining first-principles calculations and the newly developed machine learning method. In this spin Hamiltonian, we have successfully unveiled the magnetic interactions of XPS$_3$ system. We find that the AFM-ZZ ground state within FePS$_3$ monolayer is stabilized by competing ferromagnetic nearest-neighbor and antiferromagnetic third nearest-neighbor exchange interactions, and combining single-ion anisotropy. However, the often ignored nearest-neighbor biquadratic exchange is crucial interaction for the stabilization of the AFM-ZZ order within NiPS$_3$. By adopting our model, one can also accurately calculate the spin wave, which paves a way for the future experimental study of magnetic excitations in XPS$_3$ systems. We believe that the exact spin Hamiltonian discovered in this study could be widely used in understanding magnetic interactions of two-dimensional materials.

\section*{ACKNOWLEDGEMENTS}
We acknowledge financial support from the Ministry of Science and Technology of the People's Republic of China (No. 2022YFA1402901), the support from the National Natural Science Foundation of China (Grants No. 11825403, No. 12074301, No. 12004295). P. Li also acknowledge supports from the China's Postdoctoral Science Foundation funded project (Grant No. 2022M722547), and the Open Project of State Key Laboratory of Surface Physics (No. KF2022$\_$09).



\begin{thebibliography}{99}
\bibitem{37} LDA+U+SOC (FePS$_3$) or GGA+U (NiPS$_3$) calculations using the structure optimized with the FM order, and full LDA+U+SOC (FePS$_3$) or GGA+U (NiPS$_3$) calculations in which the structures are optimized with the corresponding magnetic orders.
\end{thebibliography}
\end{document}